\documentclass[11pt,a4paper]{amsart}

\usepackage[latin1]{inputenc}
\usepackage[english]{babel}
\usepackage{amsmath}

\usepackage{amssymb, mathabx}
\usepackage[numbers]{natbib}
\usepackage{graphicx}
\usepackage{braket}
\usepackage[pdftex,plainpages=false,colorlinks,hyperindex,bookmarksopen,linkcolor=red,citecolor=blue,urlcolor=blue]{hyperref}

\usepackage{mathrsfs}

\usepackage{epstopdf}

\usepackage{braket}

\bibpunct{[}{]}{;}{n}{,}{,}
\usepackage[hmargin=3cm,vmargin={3.5cm,4cm}]{geometry}

\theoremstyle{theorem}

\theoremstyle{definition}                                 %stile corsivo
\theoremstyle{definition}                           %stile roman
\theoremstyle{remark}                             %stile per osservazioni
\newtheorem*{rmk}{Remark}              %definizione ambiente osservazione

\usepackage{color}

\usepackage{mathtools,slashed}

\newcommand{\be}{\begin{eqnarray}}
\newcommand{\ee}{\end{eqnarray}}

\newcommand{\R}{\mathbb{R}}  %%%%% \R = \mathbb{R}.
\newcommand{\C}{\mathbb{C}} %%%% \C = \mathbb{C}.
\newcommand{\N}{\mathbb{N}} %%%% \K = \mathbb{K}.
\def\eg{{\it e.g. }} 
\def\ie{{\it i.e. }}

\newcommand{\wt}[1]{\widetilde{#1}}

\newcommand{\ceil}[1]{\lceil #1 \rceil}

\def\eg{{\it e.g.}\ }
\def\ie{{\it i.e.}\ }

\numberwithin{equation}{section}

\allowdisplaybreaks

\begin{document}
\title{Prabhakar-like Fractional Viscoelasticity}
	
	    \author{Andrea Giusti$^1$}
		\address{${}^\dagger$ Department of Physics $\&$ Astronomy, University of 	
    	    Bologna and INFN. Via Irnerio 46, Bologna, ITALY and 
	    	 Arnold Sommerfeld Center, Ludwig-Maximilians-Universit\"at, 
	    	 Theresienstra{\ss}e~37, 80333 M\"unchen, GERMANY.}	
 		\email{agiusti@bo.infn.it}
 		
	\author{Ivano Colombaro$^2$}
		\address{${}^1$ Department of Information and Communication Technologies, 
		Universitat Pompeu Fabra. C/Roc Boronat 138, 08018, Barcelona, SPAIN.}
		\email{ivano.colombaro@upf.edu} 
 

	\thanks{In: \textbf{Comm.~Nonlin.~Sci.~Num.~Sim.~56C~(2018) pp.~138--143
}, \textbf{DOI}: \href{http://www.sciencedirect.com/science/article/pii/S1007570417302812}{10.1016/j.cnsns.2017.08.002}}	
	
    \date  {\today}%%{January 2016}

\begin{abstract}
The aim of this paper is to present a linear viscoelastic model based on Prabhakar fractional operators. In particular, we propose a modification of the classical fractional Maxwell model, in which we replace the Caputo derivative with the Prabhakar one. Furthermore, we also discuss how to recover a formal equivalence between the new model and the known classical models of linear viscoelasticity by means of a suitable choice of the parameters in the Prabhakar derivative. Moreover, we also underline an interesting connection between the theory of Prabhakar fractional integrals and the recently introduced Caputo-Fabrizio differential operator. 
\end{abstract}

    \maketitle

%%%%%Section 1
%%%%%%%%%%

\section{Introduction}
	Various notions of generalized Mittag-Leffler function with three parameters \cite{FM-ML} have been subject to an extensive interests in the last few years. Among all the possible definitions, the one proposed by Prabhakar in \cite{PBK} seems to play a fundamental role in both mathematics and physics, see \eg \cite{Capelas, GGPT, Garrappa}. On this note, it is important to stress that this function allows for an extention of the theory of fractional integro-differential operators which is known in the literature as Prabhakar-like fractional calculus. The growing interest in this approach appears to be particularly justified by the implications that can be drawn by its application to the theory of probability and of stochastic processes \cite{PT}.
	
	In this paper, after a brief review of the main results concerning both Mittag-Leffler functions and Prabhakar calculus, we present an example of linear viscoelastic system based on the Prabhakar fractional derivative. Besides, we also present a connection between a recently developed differential operator, introduced by M.~Caputo and M.~Fabrizio in \cite{Caputo-Fabrizio}, and the theory of Prabhakar fractional integrals.

\subsection{Mittag-Leffler function and its generalizations} \label{Sec-1-1}
	As widely discussed in the literature (see \eg \cite{FM-ML, HMS, Mainardi_BOOK10, Mittag-Leffler}) the Mittag-Leffler function can be thought of as a special function that generalizes the exponential function. Indeed, this extension is obtained by means of a slight modification of the power series expansion for the exponential, precisely
	\be \label{ML-1}
	E_\alpha (z) = \sum _{k=0} ^\infty \frac{z^k}{\Gamma (\alpha \, k + 1)} \, , 
	\qquad z, \alpha \in \C, \, \texttt{Re}(\alpha) >0 \, .
	\ee
This function was first proposed by M.~G.~Mittag-Leffler in 1903 (see \cite{Mittag-Leffler}).
	
	A first generalization of the function \eqref{ML-1} was proposed by A.~Wiman in 1905 \cite{Wiman} as a two parameters function given by
	\be \label{ML-2}
	E_{\alpha , \beta} (z) = \sum _{k=0} ^\infty \frac{z^k}{\Gamma (\alpha \, k + \beta)} \, , 
	\qquad z, \alpha, \beta \in \C, \, \texttt{Re}(\alpha) >0 \, .
	\ee
It is also important to remark that both these functions are known to be entire functions of order $\rho = 1/ \alpha$ and type $\sigma = 1$.

	The important property of the functions \eqref{ML-1} and \eqref{ML-2} is that they are intrinsically intertwined with the theory of differential equations of fractional order. Indeed, it is very well known that these functions arise naturally in the solutions of this kind of differential equations, for further details see \eg \cite{FM-ML, HMS, Mainardi_BOOK10, Mainardi-1997}.

	Another main feature of the Mittag-Leffler functions with one and two parameters is that they reproduce both a purely exponential and purely power-law asymptotic behaviors. Therefore, this property makes them a fundamental tool for the study of fractional relaxation processes.

	Again, by means of the series representation, a further generalization of \eqref{ML-1} and \eqref{ML-2} was proposed by Prabhakar in 1971 \cite{PBK} 
	\be \label{PBK-function} 
	E ^\gamma _{\alpha , \beta} (z) = 
	\sum _{k=0} ^\infty \frac{(\gamma) _k}{\Gamma (\alpha \, k + \beta)} \frac{z^k}{k!} \, ,
	\ee	
	where $z \in \C$, $\alpha, \beta, \gamma \in \C$, $\texttt{Re}(\alpha) > 0$, and where $(\gamma) _k$ is the Pochhammer (rising factorial) symbol, that can also be rewritten as $ (\gamma) _k = \Gamma (\gamma + k) / \Gamma (\gamma)$.
	
	The latter is known to be an entire function of order $\rho = 1/ \texttt{Re} (\alpha)$ and type $\sigma = 1$. Moreover, it is also trivial to see that
	$$ E_\alpha (z) = E ^1 _{\alpha , 1} (z) \, , \,\,\,  E_{\alpha , \beta} (z) = E ^1 _{\alpha , \beta} (z) \, , \,\,\, 
	\exp (z) = E ^1 _{1 , 1} (z) \, .$$
	
	It is also important to recall a peculiar Laplace transform of \eqref{PBK-function} as it will turn out to be quite useful in the following. Specifically, it was shown in \cite{KSS} that
	\be \label{eq-LT-gml}
	\mathcal{L} \left\{ t^{\beta -1} \, E^{-\gamma} _{\alpha , \, \beta} ( \omega \, t^\alpha ) \right\} 
	= 
	s^{- \beta} \, \left( 1 - \omega \, s^{-\alpha} \right) ^\gamma
	\, \, ,
	\ee
	where $t \in \R$, $\alpha, \beta, \gamma, \omega \in \C$ and $\texttt{Re}(\alpha) > 0$.
	
	For further details on generalized Mittag-Leffler type functions we invite the interested reader to refer to \eg \cite{FM-ML, HMS, KSS, MG, PT}.
	
	\subsection{Prabhakar Fractional Operators} \label{Sec-1-2}
	As argued by various authors (see \eg \cite{GGPT, HMS, KSS, MG, PT}), the generalized Mittag-Leffler with three parameters allows for a generalization of the Riemann-Liouville-Caputo fractional calculus. Indeed, first let us define the Prabhakar kernel
	\be \label{P-kernel} 
	e ^\gamma _{\alpha , \beta} (\omega ; \, t) := t^{\beta - 1} \, E ^\gamma _{\alpha , \beta} (\omega t^\alpha) \, ,
	\ee
	where $t \in \R$, $\alpha, \beta, \gamma, \omega \in \C$ and $\texttt{Re}(\alpha) > 0$.
	
	Now, let $0 \leq a < t < b \leq + \infty$, then given a function $f \in L^1 \left(a, \, b\right)$ one can define the \textbf{Prabhakar integral} as
	\be \label{P-integral} 
	_a \textbf{E} ^\gamma _{\alpha , \beta , \omega} \, f (t) &\!\!:=\!\!& 
	\left( f \ast e ^\gamma _{\alpha , \beta} (\omega ; \cdot) \right) (t) \\
	&\!\!=\!\!& \int _a ^t (t - \tau ) ^{\beta - 1}  E ^\gamma _{\alpha , \beta} \left[ \omega \, (t - \tau)^\alpha \right] \, 
	f(\tau) \, d \tau \notag
	\, .
	\ee
	
	\begin{rmk}
	It is easy to see that \eqref{P-integral} reduces to the Riemann-Liouville fractional integral either as $\gamma$ goes to zero or when $\gamma$ tends to one and $\omega$ vanishes. Indeed, $e ^0 _{\alpha , \beta} (\omega ; t) = e ^1 _{\alpha , \beta} (0 ; t) = t^{\beta - 1} / \Gamma (\beta)$, which is exactly the Riemann-Liouville integral kernel.
	\end{rmk}
	
	Now, in strict analogy with the classical analysis of fractional operators, we can introduce the (regularized) Prabhakar derivative in a straightforward manner.
	
	Let $m = \ceil{\beta}$ and $f \in AC^m \left(a, \, b\right)$, then the \textbf{regularized Prabhakar derivative} \cite{D'Ovidio-Polito} reads
	\be \label{P-derivative}
	^C _a \textbf{D} ^\gamma _{\alpha , \beta , \omega} \, f (t) = 
	  {}_a \textbf{E} ^{-\gamma} _{\alpha , m - \beta , \omega} \, f^{(m)} (t)
	\ee
	where $f^{(m)} (t)$ represents the $m$th derivative of $f(t)$ and $AC^m \left(a, \, b\right)$ stands for the set of real-valued functions $f(t)$ whose derivatives are continuous up to order $m-1$ on $(a, \, b)$ and such that $f^{(m-1)} (t)$ is an absolutely continuous function.
	
	\begin{rmk}
	Here we focus our attention only on the regularized version of the Prabhakar derivative due to its intrinsic relevance in the construction of well-posed and physically reasonable initial value problem, as widely discussed by Garra et al in \cite{GGPT} and again stressed by Polito and Tomovski in \cite{PT}.
	\end{rmk}
	
\subsection{On the Caputo-Fabrizio operator} \label{Sec-1-3}
	In a recent paper \cite{Caputo-Fabrizio}, Caputo and Fabrizio have proposed a new definition for a differential operator with a non-singular kernel. Specifically, this operator is defined as
	\be \label{CF-derivative}
	^{CF} {}_a \textbf{D} ^\alpha f (t) = 
	\frac{M(\alpha)}{1 - \alpha} \int _a ^t  \exp \left[ - \frac{\alpha}{1 - \alpha} \, (t - \tau) \right] \, f' (\tau) \, d \tau \, ,
	\ee
	where $f \in L^1 \left(a, \, b \right)$, $f'(t)$ represents the first derivative of $f(t)$, $M(\alpha)$ is a normalization constant such that $M(0) = M(1) = 1$ and $0 < \alpha < 1$.
	
	Now, it is easy to see that \eqref{CF-derivative} can be related to the theory of Prabhakar integrals in a quite elegant way. Indeed, recalling that 
	\be 
	E ^1 _{1, 1} (x) = \exp (x) \, ,
	\ee
	hence
	\be 
	{}_a \textbf{E} ^1 _{1 , 1 , \omega} \,  f (t) = 
	\int _a ^t  \exp \left[ \omega \, (t - \tau) \right] \, f(\tau) \, d \tau \, .
	\ee
	Therefore,
	\be 
	{}_a \textbf{E} ^1 _{1 , 1 , \omega (\alpha) } \,  f' (t) = 
	\frac{1 - \alpha}{M(\alpha)} \, {}^{CF}  {}_a \textbf{D} ^\alpha f (t) \, ,
	\ee
	where $\omega (\alpha) = - \alpha / (1 - \alpha)$. Thus, this allows us to present a side view on the analysis of the Caputo-Fabrizio operator in terms of the Prabhakar-type fractional operators.

\subsection{On the physical meaning of the Prabhakar-type fractional operators}
	It is important to stress that Prabhakar's operators are not just some involved mathematical objects, but rather they naturally emerge from physisically relevant models for dielectric relaxation phenomena, as stressed in \cite{RG-FM-GM}, and remarked in \cite{Garrappa}. 

	In particular, it was shown by E.~Capelas de Oliveira, et al in \cite{Capelas} that the response function of the Havriliak-Negami model can be written as a specific realization of the Prabhakar kernel \eqref{P-kernel}, \ie $\beta = \alpha \gamma$, $\omega = - \lambda$, with $\lambda > 0$, and $0 < \alpha , \, \gamma < 1$. Furthermore, it has also been shown by Garrappa \cite{Garrappa} that the input-output equation of the Havriliak-Negami model, in the time domain, can be easily cast in terms of an integral equation involving the Prabhakar fractional integral. 
	
	\section{Fractional Maxwell model with Prabhakar derivatives}
	Linear viscoelasticity appears to be a preferential playground for applications of fractional calculus to realistic physical systems, see \eg \cite{IC-AG-FM-ZAMP, IC-AG-FM-Bessel, fracmax, AG-FCAA-2017, AG-FM_MECC16, AG-FM-EPJP, Mainardi_BOOK10, Mainardi-1997, Mainardi-Spada 2011}.
	
	The aim of this paper is to present an explicit realization of a simple viscoelastic model based on the Prabhakar calculus. The simplest way to proceed towards our goal is to consider the constitutive equation of the classical fractional Maxwell model in which we replace the Caputo derivative (or Riemann-Liouville, considering that there are no appreciable differences between these two formulations of the problem, as discussed in \cite{Bagley}) with the (regularized) Prabhakar one.
	
	That said, let us consider two causal functions $\sigma , \varepsilon$, respectively representing the uniaxial stress and strain for a certain system, such that $\sigma , \varepsilon \in AC^1 \left(0, \, + \infty \right)$, with $\alpha, \beta, \gamma, \omega \in \R$, $\alpha > 0$ and $0 < \beta < 1$.  Furthermore, let us consider a stress-strain relation given by
	\be \label{P-maxwell}
	\sigma (t) + a \, ^C \textbf{D} ^\gamma _{\alpha , \beta , \omega} \, \sigma (t) = 
	b \, ^C \textbf{D} ^\gamma _{\alpha , \beta , \omega} \, \varepsilon (t)
	\ee
	where $a, b \in \R$, and where $^C \textbf{D} ^\gamma _{\alpha , \beta , \omega} \equiv {}^C {}_{0+} \textbf{D} ^\gamma _{\alpha , \beta , \omega}$. 
	
	It is important to point out that here it is not mandatory to require $a, b > 0$ in order to recover physically meaningful models \cite{Hanyga STAMM04}. The reason for this broader freedom is due to the large number of parameters involved in the definition of the Prabhakar derivative. The general analysis of physically acceptable Prabhakar-like viscoelastic models is a very interesting problem per se, however a complete characterization of the latter is beyond the scope of this paper and it will therefore be tackled in a future development of the theory. 
	
	Now, recalling that the Laplace transform of the regularized Prabhakar derivative is given by (see \cite{GGPT}),
	\be 
	\mathcal{L} \left\{ ^C \textbf{D} ^\gamma _{\alpha , \beta , \omega} \, f(t) \, ; \, s \right\} =
	s^\beta \left( 1 - \omega \, s^{-\alpha} \right)^\gamma \, \wt{f} (s) 
	- s^{\beta - 1} \, \left( 1 - \omega \, s^{-\alpha} \right)^\gamma \, f(0+) \, ,
	\ee 
	where $\wt{f} (s) \equiv \mathcal{L} \left\{ f(t) \, ; \, s \right\}$, it is easy to see that \eqref{P-maxwell} turns into
	\be \label{P-Maxwell-Laplace}
	\left\{ 1 + a \, s^\beta \, \left( 1 - \omega \, s^{-\alpha} \right) ^\gamma \right\} \wt{\sigma} (s)
	=
	b \, s^\beta \, \left( 1 - \omega \, s^{-\alpha} \right) ^\gamma \, \wt{\varepsilon} (s) \, ,
	\ee
	in the Laplace domain, upon assuming $a \,  \sigma (0+) = b \, \varepsilon (0+)$. Notice that this last condition on the initial data represents a quite reasonable constraint, as explained in \cite{AG-FCAA-2017, Mainardi_BOOK10, Mainardi-Spada 2011}.
	
	From the general theory of linear viscoelasticity \cite{Mainardi_BOOK10} one has that a given linear model can be expressed in terms of two equivalent forms, namely the creep and the relaxation representations. To each of these representations we associate a material function: $G(t)$ and $J(t)$, respectively called relaxation modulus and creep compliance. It is also important to stress that each of these two functions contains all the physical information about the viscoelastic model. Concretely, in the Laplace domain, a given (linear) constitutive equation takes the following two equivalent forms \cite{Gurtin-Sternberg, Mainardi_BOOK10}
	\be \label{creep-relaxation}
	\wt{\sigma} (s) = s \, \wt{G} (s) \, \wt{\varepsilon} (s) \, , \qquad 
	s \, \wt{J} (s) \, \wt{\sigma} (s) = \wt{\varepsilon} (s) \, ,
	\ee
	if we assume some suitable conditions on the initial data (see \cite{AG-FCAA-2017, Gurtin-Sternberg, Mainardi_BOOK10, Mainardi-Spada 2011} for further details).
	
	According to \eqref{creep-relaxation}, we can now easily deduce the material functions, in the Laplace domain, just by inspection of \eqref{P-Maxwell-Laplace}, \ie
	\be 
	s \, \wt{J} (s) = \frac{a}{b} + \frac{1}{b \, s^\beta \, \left( 1 - \omega \, s^{-\alpha} \right) ^\gamma} 
	= 
	\frac{1}{s \, \wt{G} (s)} \, ,
	\ee
	and therefore
	\be \label{J-tilde}
	\wt{J} (s) 
	= 
	\frac{a}{b \, s} + \frac{1}{b \, s^{\beta + 1} \, \left( 1 - \omega \, s^{-\alpha} \right) ^\gamma} \, ,
	\ee
	\be \label{G-tilde}
	\wt{G} (s) = 
\frac{b}{a \, s} \left[1 + \frac{1}{a \, s^{\beta} \, \left(1 - \omega \, s^{-\alpha} \right)^{\gamma}} \right]^{-1} \, .
	\ee
	
	Then, taking profit of \eqref{eq-LT-gml} we can immediately invert $\wt{J} (s)$ back to the time domain, and we get 
	\be \label{J}
	J (t) = \frac{a}{b} + \frac{t^\beta}{b} \, E^\gamma _{\alpha , \, \beta +1} (\omega \, t^\alpha) \, .
	\ee
	
	Let us now focus on $G(t)$, despite the easy procedure that has lead to \eqref{J}, computing the explicit form of the relaxation modulus is rather less trivial. Luckily, we can take profit of some results and procedures discussed in \cite{GGPT}.
	
	First, from \eqref{G-tilde} it is easy to see that if  
	\be \label{eq-LT-condition}
	\left| \frac{1}{a \, s^{\beta} \, \left(1 - \omega \, s^{-\alpha} \right)^{\gamma}} \right| < 1 \, ,
	\ee
	we can then expand $\wt{G} (s)$ as an absolutely convergent power series, \ie
	\be 
	\wt{G} (s) &\!\!=\!\!& \frac{b}{a \, s} \sum _{n=0} ^\infty \left[ - \frac{1}{a \, s^{\beta} \, \left(1 - \omega \, s^{-\alpha} \right)^{\gamma}} \right] ^n
	=\\
	&\!\!=\!\!&
	\frac{b}{a} \sum _{n=0} ^\infty (- a)^{-n} \, s^{-(\beta n + 1)} \, \left(1 - \omega \, s^{-\alpha} \right)^{- \gamma n} \, .
	\ee
	Therefore, inverting the Laplace transform, taking advantage of \eqref{eq-LT-gml}, yields 
	\be \label{G}
	G(t) = \frac{b}{a} \sum _{n=0} ^\infty (- a)^{-n} \, t^{\beta n} \, E ^{\gamma n} _{\alpha , \, \beta n + 1} \left( \omega t^\alpha \right) \, ,
	\ee
	where the integration term by term is allowed by the fact that the generalized Mittag-Leffler function is an absolutely convergent series, provided that we chose a sufficiently large abscissa for the Bromwich path (see \cite{GGPT} for further details on this procedure).
	
	Now we just need to prove the convergence of the series in \eqref{G}. To do that it is sufficient to notice that \eqref{G} consists of repeated series, indeed
	\be 
	G(t) = 
	\frac{b}{a} \sum _{n=0} ^\infty (- a)^{-n} \, t^{\beta n} \, \sum _{k=0} ^\infty \frac{(\gamma n) _k}{\Gamma (\alpha \, k + \beta n + 1)} 
	\frac{\left( \omega t^\alpha \right)^k}{k!} 
	\, .
	\ee
	Then, given that the generalized Mittag-Leffler function is an entire function, in order to prove the absolute convergence of the series labelled by $n$, one just have to show that the series 
	\be 
	\sum _{n=0} ^\infty (- a)^{-n} \, t^{\beta n} \, \frac{(\gamma n) _k}{\Gamma (\alpha \, k + \beta n + 1)} \frac{\left( \omega t^\alpha \right)^k}{k!} \, ,
	\ee
	is absolutely convergent for each (fixed) $k \in \N \cup \{0\}$.
	
	If we define
	\be 
	a _n (k; t) = \frac{(- a)^{-n} \, t^{\beta n} \, (\gamma n) _k}{\Gamma (\alpha \, k + \beta n + 1)} \frac{\left( \omega t^\alpha \right)^k}{k!} =
	\frac{ (- a)^{-n} \, \Gamma (\gamma n + k)}{\Gamma (\gamma n) \, \Gamma (\alpha \, k + \beta n + 1)} \, \frac{\left( \omega t^\alpha \right)^k}{k!} \, 
	t^{\beta n} \, ,
	\ee
	and recalling the asymptotic behaviour of the rate of gamma functions (see \cite{ET, GGPT})
	$$ \frac{\Gamma \left( z + a \right)}{\Gamma \left( z + b \right)} \sim z^{a - b} \, , $$
	for $ |z| \to \infty$, $|\texttt{Arg} (z)| \leq \pi - \epsilon$, $|\texttt{Arg} (z + a)| \leq \pi - \epsilon$, $0 < \epsilon < \pi$, it is easy to see that
	\be 
	\left| \frac{a _{n+1} (k; t)}{a _n (k; t)} \right| \sim \left| \frac{t^\beta}{\beta \, a} \, \frac{1}{n} \right| \, , \qquad n \to \infty \, , \,\,\, \forall t > 0, \forall k \in \N \cup \{0\} 
	\, ,
	\ee
	hence
	\be 
	\lim _{n \to \infty} \left| \frac{a _{n+1} (k; t)}{a _n (k; t)} \right| = 0 \, , \qquad \forall t > 0, \, \forall k \in \N \cup \{0\} \, ,
	\ee
	that concludes our proof of the absolute convergence of \eqref{G}.
	
	\subsection{Connection with the classical models of linear viscoelasticity}
	Because of the large number of parameters that appear in \eqref{P-maxwell}, it is interesting to investigate if and under which conditions it is possible to recover some known results of linear viscoelasticity. 
	
	A simple way to do that is by comparing the Laplace transform of one of the material functions for the Maxwell-Prabhakar model with the corresponding material function of the classical viscoelastic models known in the literature. In this paper we will focus our attention on the creep compliance $\wt{J} (s)$, that for the Maxwell-Prabhakar model reads
	\be \label{eq-edit-sJs}
	s \, \wt{J} (s) &\!\!=\!\!& \frac{a}{b} + \frac{1}{b \, s^\beta \, \left( 1 - \omega \, s^{-\alpha} \right) ^\gamma} \\ \notag
	&\!\!=\!\!& \frac{a}{b} + \frac{1}{b \, s^{\beta - \alpha \gamma} \, \omega ^\gamma \, \left( s^{\alpha}/\omega -1 \right) ^\gamma}  \, .
	\ee

	Let us begin with the \textbf{fractional Maxwell model} of order $\nu$ (see \cite{IC-AG-FM-TWLV, Mainardi-Spada 2011}), defined in terms of a stress-strain relation given by
	\be 
	\sigma (t) + A \, ^C \textbf{D} ^\nu \, \sigma (t) = 
	B \, ^C \textbf{D} ^\nu \, \varepsilon (t) \, ,
	\ee
with $^C \textbf{D} ^\nu$ representing the Caputo derivative, and corresponding to a constitutive creep compliance, in the Laplace domain, given by
	\be
	s \, \wt{J} _M (s) = \frac{A}{B} \, \left[ 1 + \frac{1}{(\tau \, s)^\nu} \right] \, ,
	\ee
	with $A, B > 0$, $\tau ^\nu = A$ and $0 < \nu < 1$. Now, comparing $\wt{J} _M (s)$ with \eqref{eq-edit-sJs} one can infer that there are two configurations of the parameters that allow to recover the classical fractional Maxwell model from the Maxwell-Prabhakar one, precisely
	\begin{itemize}
	\item[(i)] $\gamma = 0$, $a = A$, $b = B$, $\beta = \nu$, $\omega \in \R$;
	\item[(ii)] $\gamma \in \R$, $a = A$, $b = B$, $\beta = \nu$, $\omega = 0$.
	\end{itemize}
	
	Another interesting case is the so called fractional Voigt model of order $\nu$ (see \cite{IC-AG-FM-TWLV, Mainardi-Spada 2011}), which is defined by its constitutive equation, \ie
	\be 
	\sigma (t) = M \, \varepsilon (t) + B \, ^C \textbf{D} ^\nu \, \varepsilon (t) \, ,
	\ee
 with a creep compliance, in the Laplace domain, given by
	\be
	s \, \wt{J} _V (s) = \frac{1}{M \, \left[1 + (\tau \, s)^\nu \right]} \, ,
	\ee
where $0 < \nu < 1$, $M, B > 0$ and $\tau  ^\nu = B / M$ (see \cite{Mainardi-Spada 2011}). 
As above, comparing the latter with \eqref{eq-edit-sJs} one can easily infer that the Maxwell-Prabhakar model \eqref{P-maxwell} reduces to a model which is formally equivalent to the fractional Voigt model if $\gamma = 1$, $a = 0$, $b = - B$, $\alpha = \beta = \nu$, $\omega = - M / B$.

	Finally, as a last example, let us consider the fractional Zener model of order $0 < \nu < 1$, \ie
	\be 
	\sigma (t) + A \, ^C \textbf{D} ^\nu \, \sigma (t) = M \, \varepsilon (t) + B \, ^C \textbf{D} ^\nu \, \varepsilon (t) \, ,
	\ee
	and corresponding to a creep compliance, in the Laplace domain, given by
	\be
s \, \wt{J} _Z (s) = \frac{1}{M} \, 
\frac{1 + A \, s^\nu}{1+ B \, s^\nu / M } \,.
	\ee
	with $M, A, B > 0$. 
	
	Again, in order to recover from \eqref{P-maxwell} a model which is formally equivalent to the fractional Zener model one just have to set the parameters in \eqref{eq-edit-sJs} as follows
	\begin{itemize}
	\item[(i)] $\gamma = 1$, $a = AB / (B - AM)$, $b = B^2  / (B - AM)$, $\alpha = \beta = \nu$, $\omega = - M / B$;
	\item[(ii)] $\gamma = 1$, $a = B / (AM - B)$, $b = M \, a$, $\alpha = \nu$, $\beta = 0$, $\omega = - M / B$.
	\end{itemize}

\section{Conclusions}
	In this paper, after an introduction about the Mittag-Leffler function and its generalizations, we provided a brief overview on Prabhakar fractional operators, that allows for a generalization of the Riemann-Liouville-Caputo fractional calculus based on the three-parameters Mittag-Leffler function. 

	Thus, we developed an example of a Prabhakar-like fractional viscoelasticity, discussing the fractional Maxwell model with Prabhakar derivatives. We then analysed the connection between the classical fractional models (Maxwell, Voigt and Zener) and the new Prabhakar-Maxwell model.

	Furthermore, in the introductory section we also pointed out an interesting formulation of the so called Caputo-Fabrizio derivative in terms of a particular Prabhakar fractional integral. 
	
\section*{Acknowledgments}
	The authors acknowledge the anonymous reviewers for the constructive comments and suggestions which have helped to improve the manuscript significantly.
	
	The work of the authors has been carried out in the framework of the activities of the National Group of Mathematical Physics (GNFM, INdAM).	
	
	Moreover, the work of A.G. has been partially supported by \textit{GNFM/INdAM Young Researchers Project} 2017 ``Analysis of Complex Biological Systems''.
	
%%%%%% Bibliography

\end{document}